# Low-loss mode converter for coupling light into slotted photonic crystal waveguide

Xingyu Zhang*,a, Harish Subbaramanb, Amir Hosseinib, Zeyu Pana, Hai Yana, Chi-jui Chunga, and Ray T. Chen*,a,b

a University of Texas at Austin, 10100 Burnet Rd, MER 160, Austin, TX 78758, USA;
b Omega Optics, Inc., 8500 Shoal Creek Blvd, Austin, TX 78757, USA.

## ABSTRACT

We design, fabricate and experimentally demonstrate a highly efficient adiabatic mode converter for coupling light into a silicon slot waveguide with a slot width as large as 320nm. This strip-to-slot mode converter is optimized to provide a measured insertion loss as low as 0.08dB. Our mode converter provides 0.1dB lower loss compared to a conventional V-shape mode converter. This mode converter is used to couple light into and out of a 320nm slot photonic crystal waveguide, and it is experimentally shown to improve the coupling efficiency up to 3.5dB compared to the V-shape mode converter, over the slow-light wavelength region.

**Keywords:** guided waves, mode converter, photonic crystal waveguides, silicon photonics, waveguides.

## 1. INTRODUCTION

In slot photonic crystal waveguides (PCWs) [1], the strong optical confinement in the slot filled with a low index material, such as air or organic polymers [2-4], is combined with the enhanced light-matter interaction provided by a slow-light structure [5, 6] for improving the device performance and miniaturizing device size. Specifically, silicon slot PCWs infiltrated with electro-optic (EO) active polymers have shown to enable high performance EO modulators [7-9], optical interconnects [4, 10], and photonic sensors [11-13]. For example, We have recently demonstrated an EO polymer infiltrated silicon slot PCW Mach–Zehnder interferometer (MZI) modulator with a switching voltage of 0.94V and an interaction length of 300 μm, corresponding to a record-high effective in-device EO coefficient ($r_{33}$) of 1230pm/V due to the combined effects of large EO polymer $r_{33}$ and slow-light enhancement [14]. In comparison, in Ref [15] a non-slow-light/non-resonant MZI modulator based on EO polymer refilled silicon slot waveguide has an large interaction length of 1.5mm, but the measured in-device $r_{33}$ is only 15pm/V. For these EO polymer based devices, the EO polymer needs to be poled under a DC electric field, so that the Pockels effect can be produced from the non-centrosymmetric alignment of the guest chromophores doped in the host amorphous polymers [16-20]. In this EO polymer poling process, the leakage current due to the charge injection through the silicon/polymer interface is known to be detrimental to the poling efficiency [21], especially for narrow slot widths ($S_w$) < 200 nm. Widening the slot has been so far the only successful approach to reduce the leakage current and improve the poling efficiency [22, 23]. It has been demonstrated that a slot width ($S_w$) as large as 320nm can significantly suppress the leakage current in the poling process and achieve an EO coefficient at the same level as that of the poled thin films of EO polymer, which is over two orders of magnitude larger compared to that in the narrow slot ($S_w$~75nm), while still achieving high optical confinement of the fundamental mode in this wide slot [22, 23]. In addition to increasing EO polymer poling efficiency for guest-host type EO polymer materials, there are also some other benefits of using large slot width as below. It was demonstrated that the poling-induced optical loss is reduced by the reduction of leakage current through large slot [24]. And also, different from typical slot widths of 100~120nm in conventional slot waveguides [25], widening the slot width to 320nm also reduces the slot capacitance, enabling higher RF bandwidth [26] and lower energy consumption [27]. Additionally, the wider slot provides other benefits such as relaxed fabrication requirement and easier infiltration of cladding material.

*xzhang@utexas.edu; phone 1 512-471-4349; fax 1 512 471-8575
*raychen@uts.cc.utexas.edu; phone 1 512-471-7035; fax 1 512 471-8575

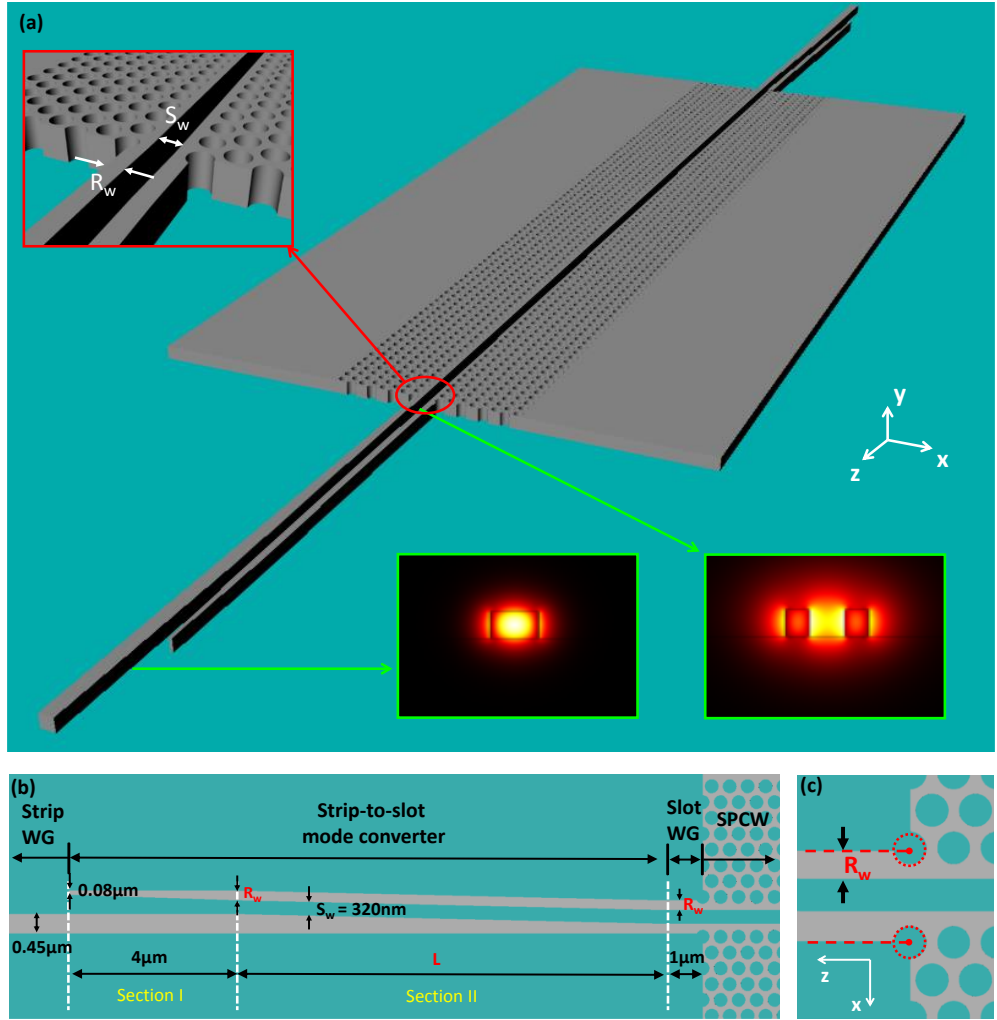

Fig. 1. (a) Schematic of our mode converter used for coupling light between a strip waveguide and a slot PCW on an SOI substrate. The top inset shows a magnified image of the coupling interface between the slot waveguide and the slot PCW. The bottom insets show the cross-sectional fundamental TE mode profile of the strip waveguide and the slot waveguide, respectively. (b) Top view of the mode converter between the strip waveguide and the slot PCW, consisting of two linearly tapered sections. Length of Sections I is fixed at 4 μm, and the length of Section II is optimized to achieve highest conversion efficiency. (c) Top view of magnified image of the coupling interface between the slot waveguide and the slot PCW. $S_w$: slot width; $R_w$: rail width; WG: waveguide; SPCW: slot photonic crystal waveguide; L: length of section II of the mode converter.

Despite the high EO polymer performance in wide slot waveguides, efficient coupling between a strip waveguide and a slot waveguide is challenging due to the large mode mismatch, as shown in the bottom insets of Fig. 1 (a) (fundamental TE mode). One common type of strip-to-slot mode converter is a V-shape mode converter [8, 28, 29]. We previously used this V-shape mode converter for coupling light from a strip waveguide into the 320nm slot PCW [22, 23]. However, the non-zero width of the fabricated tip due to lithography limitation leads to a discontinuity at the center of the mode profile, causing the total insertion loss to be as high as 23dB in which each mode converter attributes to a ~1dB insertion loss [22]. To address this problem, in this paper, we explore a new type of adiabatic mode converter to couple light from a single mode strip waveguide into a wide slot PCW, as shown in Figs. 1 (a) and (b). The mode converter consists of two linearly tapered sections, and the specific profile and dimensions are given in Fig. 1 (b). This type of adiabatic mode converter has been used for conventional narrow slot waveguides with $S_w$<130nm [30-32], and insertion losses <0.04dB in a strip-loaded slot waveguide were demonstrated [32]. However, until now, adiabatic mode converters for larger slot widths (e.g. $S_w$~320nm) have not been reported. It is to be noted that the 320nm slot waveguide here is leaky waveguide, and it needs to be specially designed to minimize the optical power leakage and get the light into/out of the PCW as soon as possible. For this reason, the slot waveguide section is designed to be as short as 1 μm as shown in Fig. 1 (b), and the rail width ($R_w$) is optimized to reduce optical loss. Moreover, contrary to conventional design rules, wherein the outer edge of the slot waveguide rails terminate at the center of holes in the first adjacent rows of the slot PCW [1, 22, 23, 29, 33-37], as shown in Fig. 1 (c), we find that if the termination is not at the center of the hole, very good coupling efficiency can still be achieved. In this work, we design, fabricate and characterize a highly efficient silicon strip-to-slot mode converter for coupling light into a 320nm slot waveguide with optical loss below 0.08dB. In addition, our adiabatic mode converter and V-shape mode converters are fabricated on the same chip, and the measured loss shows that our adiabatic mode converter has a 0.1dB lower loss compared to V-shape mode converter. Furthermore, we want to emphasize that we not only optimize a strip-to-slot mode converter for slot waveguide with $S_w$=320nm, but also demonstrate the use of this mode converter for efficiently coupling light into and out of a 300 μm-long EO polymer refilled slot PCW with the same slot width (Figs. 1 (a) and (b)). A clear band gap with about 25dB extinction ratio is observed, and an improvement of 3.5dB in coupling efficiency within the slow-light wavelength region compared to the V-shape mode converter is demonstrated. This highly efficient light coupling into wide slot PCWs, combined with the improved poling efficiency of the electro-optic (EO) polymer in wide slot PCWs [22, 23], provides tremendous advantages for several promising applications, including photonic integrated circuits, optical interconnects, EO modulation, and electromagnetic field sensing.

## 2. OPTIMIZATION OF ADIABATIC MODE CONVERTER FOR WIDE SLOT WAVEGUIDE

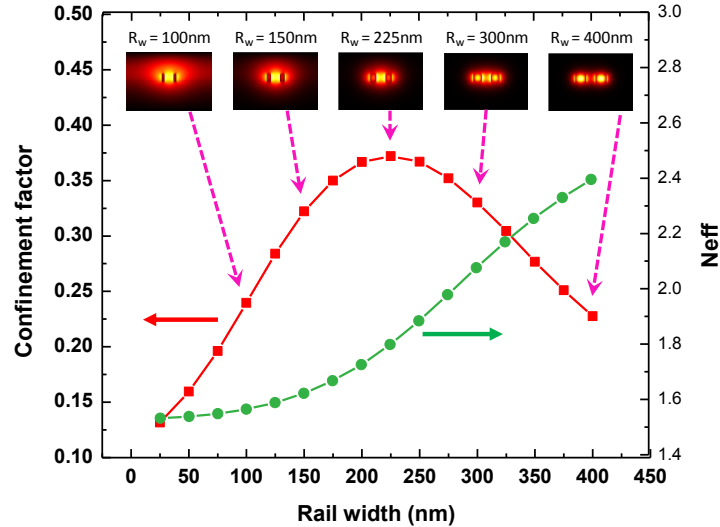

Fig. 2. Confinement factor within the slot (red curve marked with squares) and neff (green curve marked with circles) plotted as a function of rail width (Rw), overlaid with the cross-sectional fundamental TE mode profiles for different Rw. The slot width (Sw) is 320nm, and the wavelength is 1550nm.

The waveguides in this work are designed on a silicon-on-insulator (SOI) substrate with top silicon thickness of 250nm and buried oxide thickness of 3 μm. A slot waveguide is designed and used as an input for a designed slot PCW with the same $S_w$, as shown in Figs. 1 (a) and (b). The input and output strip waveguides are connected to the slot waveguides using adiabatic strip-to-slot waveguide mode converters. The slot PCW and mode converters are covered with an EO polymer cladding with refractive index of 1.63 at 1550nm wavelength. Subwavelength grating couplers are used for coupling light between the strip waveguides and single mode fibers [38, 39]. As can be seen from Fig. 1 (b), a 1 μm-long slot waveguide connects the mode converter and the slot PCW. For both the slot waveguide and the slot PCW, most electric field is confined inside the slot region. Good optical mode confinement in the slot waveguide plays an important role in increasing the coupling efficiency from the slot waveguide to slot PCW; therefore, our work starts with the optimization of this slot waveguide section. The $S_w$ is fixed at 320nm [9, 10, 23], and the rail width ($R_w$), as shown in Figs. 1 (a) and (b), of the slot waveguide is optimized for maximum mode confinement. The cross-sectional fundamental TE mode profile and the effective refractive index ($n_{eff}$) of the slot waveguide at the wavelength of 1550nm are simulated using COMSOL Multiphysics. Correspondingly, the confinement factor, defined as the overlap integral of the optical mode profile with the slot, whose mathematical expression can be found in Ref [40, 41], is also calculated. Fig. 2 shows the calculated confinement factor and $n_{eff}$ plotted as a function of $R_w$, indicating the largest confinement factor of 38% is achieved at $R_w$=225nm. In comparison, It can be seen that the conventional design with slot waveguide rails terminating at the center of holes in the slot PCW interface, for example, at $R_w$=300nm in Fig. 1(c), has a smaller confinement factor of 33%. In addition, compared to a slot waveguide with narrow $S_w$=100nm, in which a maximum confinement factor of ~42% can be achieved [40], the wider slot waveguide has a slightly lower confinement factor, but provides other advantages such as better manufacturability, better EO polymer filling, and higher EO polymer poling efficiency, which in turn provides a substantially larger EO coefficient after poling [22].

Utilizing the optimized slot waveguide, we next investigate how the length of the mode converter affects the optical loss. The mode converter consists of two linearly tapered sections, as shown in Fig 1 (b). Section I does not affect the performance of the mode converter significantly because most optical power is still confined in the 450nm-wide strip waveguide [32], so this section is fixed to be 4 μm in this work. The length of Section II, L, is critical and is optimized for achieving low enough optical loss. The $S_w$ along section II is constant and fixed to be 320nm. L is tuned from 5 μm to 30 μm, and the corresponding optical loss is simulated using FIMMWAVE. The results are shown as a blue curve in Fig. 3 (c). Next, to verify the simulation, we fabricate mode converter pairs with optimized $S_w$ of 225nm but with L varying from 5 μm to 30 μm. Test structures with different numbers of mode converters (2, 4 and 8) of varying lengths (L) connected in series are fabricated using e-beam lithography and reactive ion etching (RIE) on an SOI substrate. The total length of the strip waveguides is kept constant, so that the extracted mode converter loss is not affected by the strip waveguides. Fig. 3 (a) shows some SEM images of cascaded pairs of fabricated adiabatic mode converters with L=5 μm, 15 μm, 20 μm, and 30 μm, respectively. Then, the fabricated mode converters are covered with EO polymer as claddings using spin coating method [42, 43]. In order to test the devices, TE-polarized light from a tunable laser at 1550nm is coupled into and out of the device utilizing a grating coupler setup [38, 39]. The output optical power is measured using an optical spectrum analyzer (OSA). The measured total insertion loss of the waveguides at 1550nm (including coupling loss on gratings, propagation loss on strip waveguides, and transition loss on mode converters) for different L as a function of the total number of mode converters is plotted in Fig. 3 (b). Each measurement data point in the plot is an averaged value from three groups of identical samples. The measured optical loss per mode converter, indicated by the slope of the linear regression lines of the measured data, is extracted and plotted in Fig. 3 (c), where error bars indicate variation errors of data in the three groups of measurements. It can be seen that the measured mode converter loss decreases as the mode converter length increases. For L>25 μm, the measured mode converter loss is < 0.1dB. It can also be noticed that the variation in the measured losses becomes smaller as the length of the mode converter becomes larger. Therefore, the mode converter length is finally chosen to be 30 μm, which is the point of diminishing returns in Fig. 3 (c). The measured results match the simulation results pretty well, as shown in Fig. 3 (c). These measured losses are reproducible, and the deviations around the mean value are mainly caused due to fabrication induced errors.

Additionally, the optical bandwidth of the mode converter is investigated. The optical loss of a single adiabatic mode converter is simulated by FIMMWAVE over a wavelength range from 1520 to 1580, as shown by the blue curve in Fig. 3 (d). The transmission spectrum of 4 pairs of adiabatic mode converters (total number of 8) are measured and normalized. Then the measured loss per mode converter can be obtained by dividing this total loss by 8, as shown by the red curve in Fig. 3 (d). It can be seen that the simulation and testing results agree well with each other, indicating that our adiabatic mode converter can provide a wide low-dispersion operation.

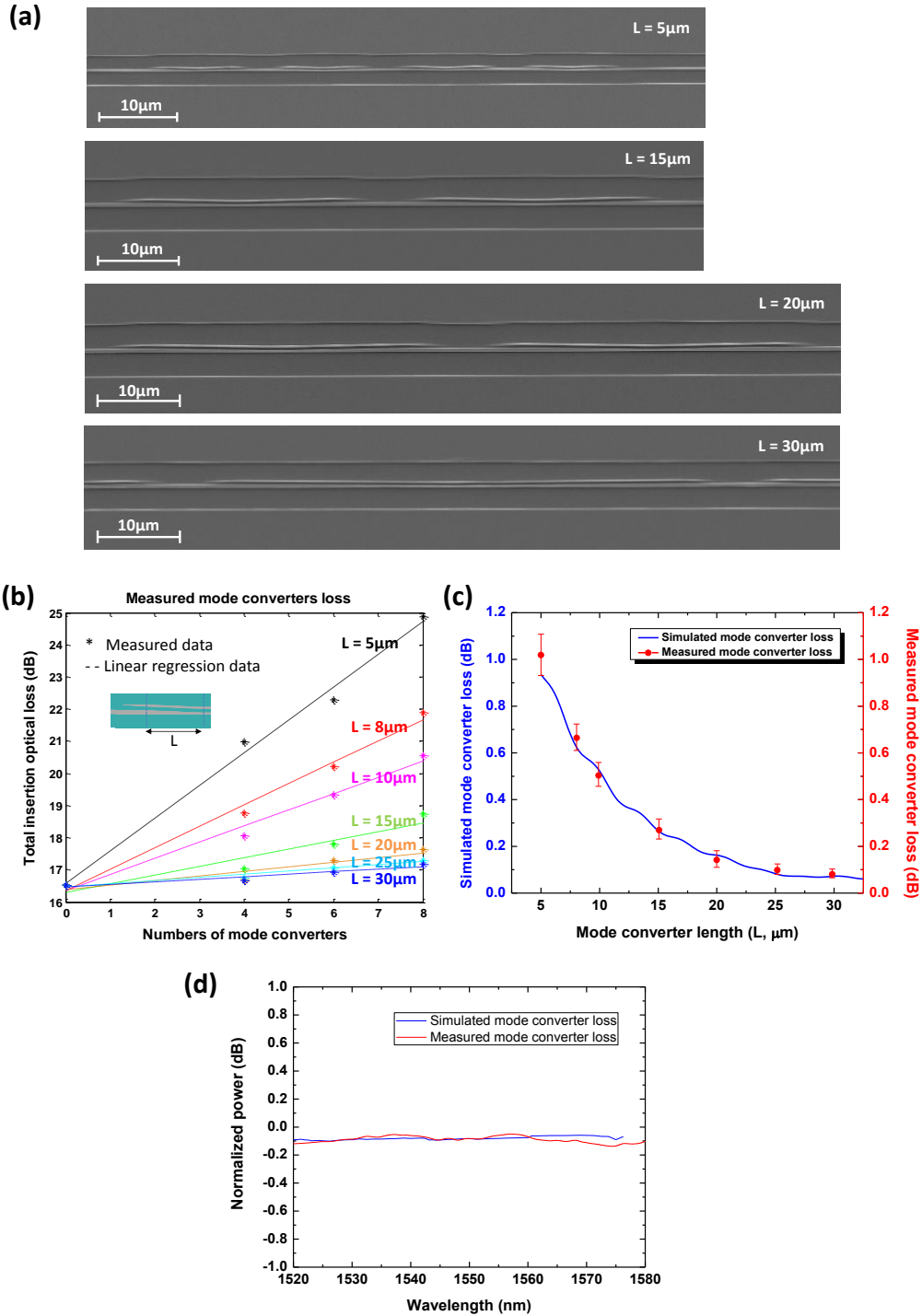

Fig. 3. (a) SEM images of fabricated test structures consisting of cascaded pairs of mode converters with L=5 μm, 15 μm, 20 μm and 30 μm, respectively. Note: here polymer claddings are not shown for better visualization. (b) Measured insertion loss indicated by dots for three fabricated samples as a function of number of mode converters in the measured arm. The loss is measured at 1550nm. (c) Simulated (blue curve) and measured (red dots) mode converter loss v.s. mode converter length. The error bars indicate the variation range of data in three groups of measurements. (d) Normalized transmission spectrum of one adiabatic mode converter. The simulation results are from FIMMWAVE simulation of a single mode converter, and the testing results are from the measured normalized transmission spectrum of 8 mode converters divided by 8.

## 3. COMPARISON BETWEEN ADIABATIC MODE CONVERTER AND V-SHAPE MODE CONVERTER

Next, we compare the performance of our optimized adiabatic mode converter with the conventional V-shaped mode converter [8, 23, 28, 29]. Both these types of converters have been explored by various groups [8, 23, 28-32]. The single-mode strip waveguide at the input has a width of 450nm, and the slot waveguide has $S_w$ of 320nm and $R_w$ of 225nm. Figs. 4 (a) and (b) show the simulated fundamental TE mode profiles (cross-sectional view), $n_{eff}$ transitions, and the propagating mode (top view, normalized real part of $E_x$ calculated by three-dimensional finite-difference time-domain (FDTD) method in RSoft) along the propagation direction for these two types of mode converters, respectively. All the simulations are done at the wavelength of 1550nm. It can be seen that our mode converter results in a smooth transformation of mode profiles and an adiabatic transition of $n_{eff}$ from a strip mode to a slot mode, as shown in Fig. 4 (a). In comparison, the V-shape mode converter has been simulated with a non-zero tip width (~80nm-wide) due to practical lithography limitations. Due to a discontinuity in the mode field distribution at the non-zero tip, an abrupt change of $n_{eff}$ occurs, as shown in Fig. 4 (b), resulting in additional optical scattering loss. Note that although an 80nm-wide tip in Section I of our adiabatic mode converter is also included in the simulation, no significant scattering is observed at this non-zero tip based on simulation results. This is because most of the electric field is confined in the 450nm-wide strip waveguide at the cross section where the non-zero tip appears, as shown in Fig. 4 (a). In addition, along our strip-to-slot mode converter, a possible second-order slot mode is suppressed due to the asymmetric slot waveguide geometry of the transition region, so the power is more efficiently coupled to the fundamental mode of the slot waveguide [44, 45].

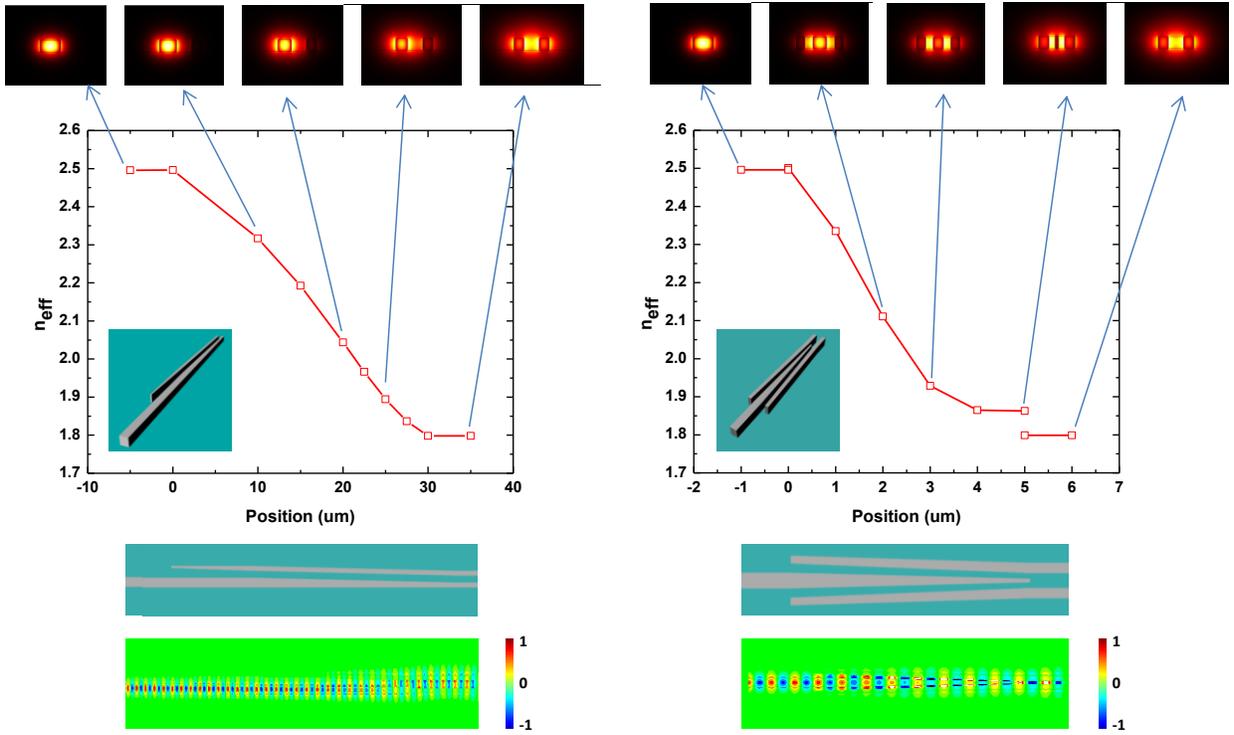

Fig. 4. The simulated $n_{eff}$ transition along (a) our mode converter and (b) a conventional V-shape mode converter, respectively, overlaid with mode profiles transformation (cross-sectional view) and FDTD simulation of mode propagation (top view), at the wavelength of 1550nm. For our adiabatic mode converter, $S_w$=320nm, $R_w$=225nm, L=30 μm. For V-shape mode converter, $S_w$=320nm, $R_w$=225nm, L=5 μm.

For experimental demonstration, a series of our strip-to-slot mode converters (L=30 μm, S1: $R_w$=225nm) together with conventional V-shape mode converters (5 μm-long, V1: $R_w$=225nm, V2: $R_w$=300nm) are fabricated on the same chip, and the insertion losses at 1550nm of these mode converters are measured and compared. Additionally, another type of mode converter (S2, $R_w$=225nm) used in Ref [11], with the same length, is also fabricated on the same chip and tested. S2 has a 4 μm-long linearly tapered section I and 10 μm-long section II similar to S1 but both with a narrow slot width of 120nm, and then a 20 μm-long section III with slot width linearly tapered from 120nm to 320nm. SEM images of mode converters S1, S2, V1, and V2 are shown in Figs 5 (a)-(d), respectively. Note that for S2, as shown in Fig. 5 (b), the strip waveguide was converted to a slot waveguide with $S_w$ of 120nm similar to that in Ref [32], and then tapered to slot waveguide with $S_w$ of 320nm and $R_w$ of 225nm. Also note that the only difference between V1 and V2 is that V1 uses an $R_w$ of 225nm (optimized), while V2 uses an $R_w$ of 300nm (un-optimized). Fig. 5 (e) shows the measured losses for these mode converters. The optical loss per mode converter can be estimated by the slope of the linear regression lines of the measured data. It can be clearly seen that our optimized mode converter (S1) has a loss of 0.080dB, which is at least 0.1dB smaller than those of V-shape mode converters (V1: 0.182dB; V2: 0.981dB). And also, the measured loss of S2 (0.075dB) is quite close to that of S1 (0.080dB) with the same length. Furthermore, from the comparison of the loss of V1 and V2 one can tell that the optimized $R_w$ (225nm) gives smaller loss (0.182dB) than that (0.981dB) of the un-optimized $R_w$ (300nm), and an improvement of about 0.8dB is achieved using the optimized $R_w$.

Note that the length of V-shape mode converter used here is only 5 μm which is a length commonly used in some works [8, 28, 29], but the loss of the $S_w$-optimized V-shape mode converter (V2) at 5 μm is not significantly different for loss at 30 μm, since no matter what length of V shape mode converter is used, the sudden discontinuity at the practical tip size still causes a high insertion loss. Therefore, increasing the length of the V shape mode converter does not provide any additional decrease in the insertion loss, as shown by the green curve in Fig. 6. The slight increase of loss is due to the increased mismatch at the sudden transition point (non-zero tip); whereas for the adiabatic converter, the longer length provides a greater reduction in the insertion loss. Theoretically, for L > 30 micron, the loss can be even lower as can be seen in Fig. 6.

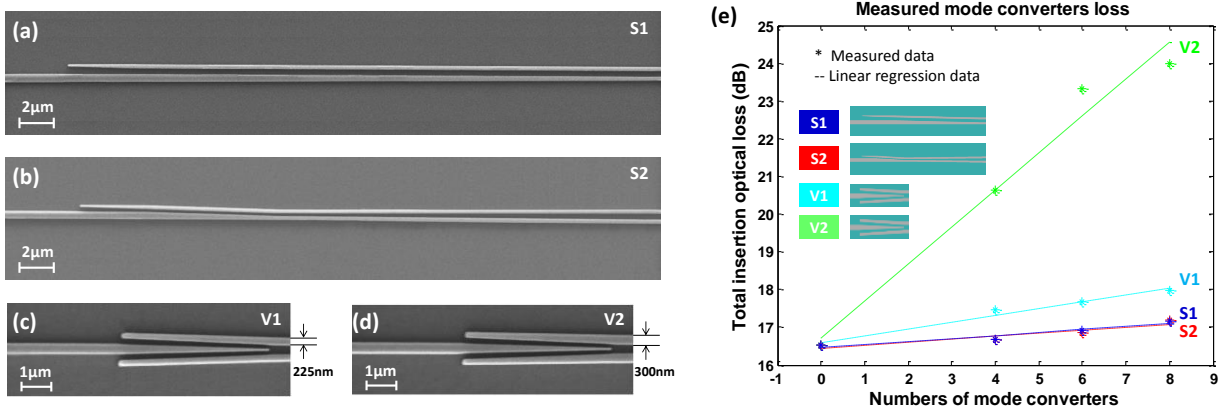

Fig. 5. SEM images of (a) our adiabatic mode converter (S1), (b) mode converter (S2) as presented in Ref [11], (c) V-shape mode converter with $R_w$=225nm (V1), and (d) V-shape mode converter with $R_w$=300nm (V2). The $S_w$=320nm for all four mode converters. L=30 μm for S1 and S2, and L=5 μm for V1 and V2. Note: here polymer claddings in (a)-(d) are not shown for better visualization. (e) Comparison of measured loss of our mode converter and conventional V-shape mode converter at 1550nm. S1: loss=0.080dB; S2: loss=0.075dB; V1: loss=0.182 dB; V2: loss=0.981dB.

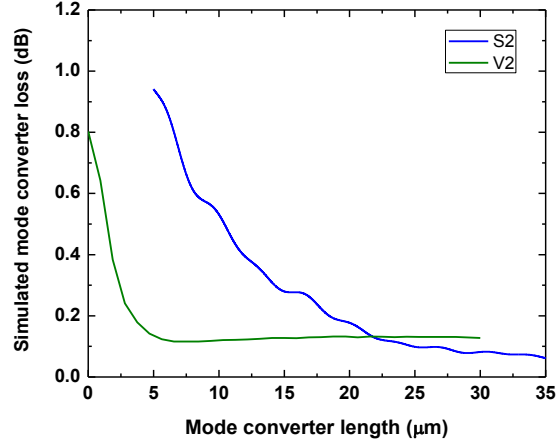

Fig. 6. Simulated loss of mode converter S2 and V2, as a function of mode converter length at the wavelength of 1550nm.

## 4. COUPLING LIGHT INTO WIDE SLOT PCW USING OPTIMIZED ADIABATIC MODE CONVERTER

Furthermore, we investigate the coupling efficiency into slot PCW at the mode converter/slot-PCW interface. For convenience, researchers previously aligned the outer edge of the rails of the slot waveguide to the center of holes in the first adjacent rows, [1, 22, 23, 29, 33-37] as shown in Fig. 1 (c). For example, in Ref [23] the slot PCW has the same structure but an un-optimized $R_w$ of 300nm is used, corresponding to a confinement factor of 33% which can be seen from Fig. 2. However, by changing the $R_w$ to 225nm (as in our optimized design) one can achieve the highest optical confinement factor of 38% in the slot, as shown in Fig. 2, with similar coupling efficiency to slot PCW.

Finally, in order to demonstrate that our optimized adiabatic mode converter (final slot rail width, $R_w$=225nm, and Section II length, L=30μm) can enable efficient light coupling between a strip waveguide and a slot PCW, a 300μm-long EO polymer infiltrated slot PCW with $S_w$=320nm (the same as the one used in Ref [23]) with our mode converter (S1) is fabricated, as shown in Fig. 7 (a), and characterized. As a comparison, the same slot PCW with the V-shape mode converter (V1) is also fabricated on the same chip, as shown in Fig. 7 (b). PCW tapers [36] are used to connect the fast-light slot waveguide with the slow-light PCW section [group index ($n_g$) of 20.4], so that the group index is gradually changed from the interface with the slot waveguide to the interface with the high $n_g$ PCW. In order to test the devices, TE polarized light from a broadband amplified spontaneous emission (ASE) source is coupled into and out of the device utilizing grating coupling setup. The optical transmission spectrum is measured by the OSA and then normalized to that of a reference strip waveguide. As shown in Fig. 7 (c), a clear band gap with more than 25dB contrast is observed in the normalized transmission spectrum of the slot PCW with our adiabatic mode converter, indicating that our optimized mode converter enables efficient coupling into the slow-light slot PCW. In comparison, using the V-shape mode converter, the band gap has a ~2dB lower contrast in the normalized transmission spectrum. Note that the Fabry-Perot reflections observed in Fig. 7 (c) are due to the PCW structure, instead of the mode converter, because it has been demonstrated in Fig. 3 (d) that the mode converter provides a flat spectrum over a wide wavelength range. This statement can also be proved from the observation that this Fabry-Perot reflections appears on both the spectra using our adiabatic mode converter and using V-shape mode converter and that no additional oscillations are introduced comparing the two. The inset of Fig. 7 (c) shows a magnified portion of the transmission spectrum in the slow-light wavelength region. The total insertion loss in the slow-light wavelength region is lower using our adiabatic mode converter compared to that using the V-shape mode converter, with a maximum loss difference of up to 3.5dB at 1560nm.

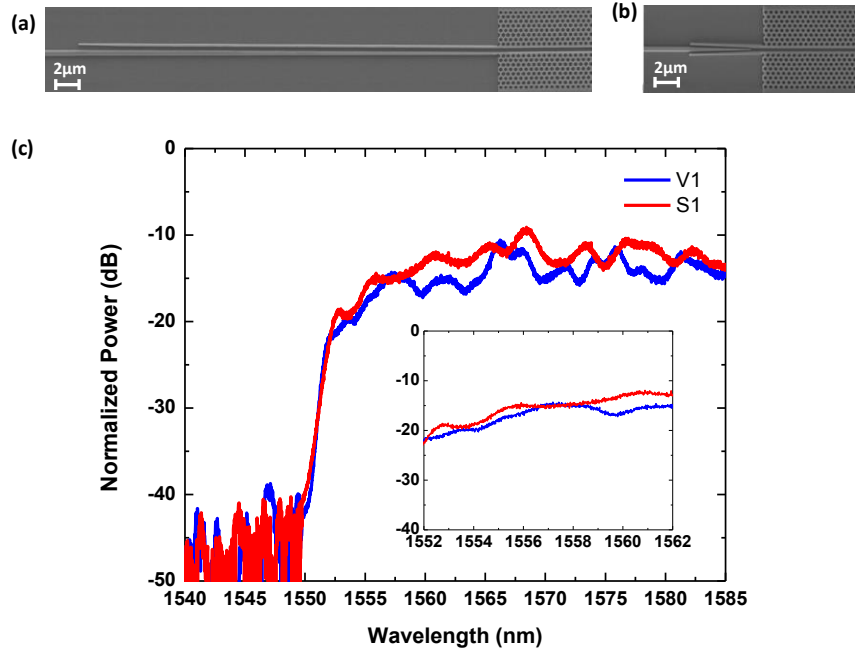

Fig. 7. SEM images of (a) our adiabatic mode converter (S1), (b) mode converter (S2) as presented in Ref [11], (c) V-shape mode converter with $R_w$=225nm (V1), and (d) V-shape mode converter with $R_w$=300nm (V2). The $S_w$=320nm for all four mode converters. L=30 μm for S1 and S2, and L=5 μm for V1 and V2. Note: here polymer claddings in (a)-(d) are not shown for better visualization. (e) Comparison of measured loss of our mode converter and conventional V-shape mode converter at 1550nm. S1: loss=0.080dB; S2: loss=0.075dB; V1: loss=0.182 dB; V2: loss=0.981dB.

## 5. CONCLUSION

In conclusion, we demonstrate a mode converter that achieves highly efficient coupling from a strip waveguide to a 320nm slot waveguide. The rail width ($R_w$) of the slot waveguide section is optimized to 225nm, yielding an optimized mode converter length of 30 μm. The measured insertion loss of the optimized mode converter is below 0.08dB at 1550nm. The optimized $R_w$ of 225nm provides a loss improvement of about 0.8dB, compared to conventional designs that require $R_w$ to be 300nm. And also, our adiabatic mode converter is demonstrated to provide a wide low-dispersion operation over a wide optical wavelength range. In addition, a comparison between our adiabatic mode converter and a conventional V-shape mode converter is presented, and an improvement of 0.1dB in loss is demonstrated for the adiabatic mode converter. Finally, in addition to coupling light between a strip waveguide and a 320nm-wide slot waveguide, our adiabatic mode converter is also used to couple light into and out of a 320nm-wide EO-polymer refilled slot PCW. We experimentally demonstrate that our mode converter provides up to 3.5dB improvement in coupling efficiency compared to the V-shape mode converter in the slow-light wavelength region of the slot PCW. This adiabatic mode converter has the advantages of low loss, easy manufacturability and large fabrication tolerance. In our future work, the loss of our mode converter can be further improved by replacing linear tapered sections by logarithmic taper profiles [32]. Furthermore, the idea of this work can be generalized and extended to the research on other slot waveguides or slot PCW structures refilled with new high-performance EO active materials, such as binary-chromophore organic glass (BCOG), consisting of shape-engineered spherical dendritic and rod-shaped dipolar chromophores, which has recently been demonstrated with an in-device EO coefficient of 230pm/V [41]. Highly efficient coupling into wide slot waveguides, combined with the improved poling efficiency of the EO active materials in wide slots, provides tremendous advantages for several promising applications, including photonic integrated circuits [46-49], optical interconnects [50-52], EO modulation [27, 53-60], and electromagnetic field detection [61-64].

## ACKNOWLEDGEMENT

The authors would like to acknowledge the Air Force Research Laboratory (AFRL) for supporting this work under the Small Business Technology Transfer Research (STTR) program (Grant No. FA650-12-M-5131) monitored by Drs. Rob Nelson and Charles Lee.

## REFERENCES

[1] A. Di Falco, L. O'Faolain, and T. Krauss, "Dispersion control and slow light in slotted photonic crystal waveguides," Applied Physics Letters, 92(8), 083501 (2008).
[2] Q. Xu, V. R. Almeida, R. R. Panepucci, and M. Lipson, "Experimental demonstration of guiding and confining light in nanometer-size low-refractive-index material," Optics letters, 29(14), 1626-1628 (2004).
[3] V. R. Almeida, Q. Xu, C. A. Barrios, and M. Lipson, "Guiding and confining light in void nanostructure," Optics letters, 29(11), 1209-1211 (2004).
[4] C. Koos, P. Vorreau, T. Vallaitis, P. Dumon, W. Bogaerts, R. Baets, B. Esembeson, I. Biaggio, T. Michinobu, and F. Diederich, "All-optical high-speed signal processing with silicon–organic hybrid slot waveguides," Nature Photonics, 3(4), 216-219 (2009).
[5] Y. A. Vlasov, M. O'Boyle, H. F. Hamann, and S. J. McNab, "Active control of slow light on a chip with photonic crystal waveguides," Nature, 438(7064), 65-69 (2005).
[6] Y. Jiang, W. Jiang, L. Gu, X. Chen, and R. T. Chen, "80-micron interaction length silicon photonic crystal waveguide modulator," Applied Physics Letters, 87(22), 221105 (2005).
[7] J.-M. Brosi, C. Koos, L. C. Andreani, M. Waldow, J. Leuthold, and W. Freude, "High-speed low-voltage electro-optic modulator with a polymer-infiltrated silicon photonic crystal waveguide," Optics Express, 16(6), 4177-4191 (2008).
[8] J. H. Wülbern, J. Hampe, A. Petrov, M. Eich, J. Luo, A. K.-Y. Jen, A. Di Falco, T. F. Krauss, and J. Bruns, "Electro-optic modulation in slotted resonant photonic crystal heterostructures," Applied Physics Letters, 94(24), 241107 (2009).
[9] X. Zhang, A. Hosseini, C.-y. Lin, J. Luo, A. K. Jen, and R. T. Chen, "Demonstration of Effective In-device r33 over 1000 pmV in Electro-optic Polymer Refilled Silicon Slot Photonic Crystal Waveguide Modulator," in CLEO: Science and Innovations, 2013, p. CTu2F. 6.
[10] X. Zhang, A. Hosseini, X. Lin, H. Subbaraman, and R. T. Chen, "Polymer-based Hybrid Integrated Photonic Devices for Silicon On-chip Modulation and Board-level Optical Interconnects," IEEE Journal of Selected Topics in Quantum Electronics, 19(6), 196-210 (2013).
[11] X. Zhang, A. Hosseini, H. Subbaraman, S. Wang, Q. Zhan, J. Luo, A. Jen, and R. Chen, "Integrated Photonic Electromagnetic Field Sensor Based on Broadband Bowtie Antenna Coupled Silicon Organic Hybrid Modulator," Lightwave Technology, Journal of, PP(99), 1-1 (2014).
[12] X. Zhang, A. Hosseini, X. Xu, S. Wang, Q. Zhan, Y. Zou, S. Chakravarty, and R. T. Chen, "Electric field sensor based on electro-optic polymer refilled silicon slot photonic crystal waveguide coupled with bowtie antenna," in SPIE Photonic West 2013: Terahertz, RF, Millimeter, and Submillimeter-Wave Technology and Applications VI, 2013, p. 862418.
[13] S. Lin, J. Hu, L. Kimerling, and K. Crozier, "Design of nanoslotted photonic crystal waveguide cavities for single nanoparticle trapping and detection," Optics letters, 34(21), 3451-3453 (2009).
[14] X. Zhang, A. Hosseini, H. Subbaraman, J. Luo, A. K.-Y. Jen, R. L. Nelson, and R. T. Chen, "Ultra-performance Optical Modulator Based on Electro-optic Polymer Infiltrated Silicon Slot Photonic Crystal Waveguide" (Under review)
[15] R. Palmer, A. Luca, D. Korn, P. Schindler, M. Baier, J. Bolten, T. Wahlbrink, M. Waldow, R. Dinu, and W.

Freude, "Low power mach-zehnder modulator in silicon-organic hybrid technology," Photonics Technology Letters, IEEE, 25(13), (2013).
[16] R. Blum, M. Sprave, J. Sablotny, and M. Eich, "High-electric-field poling of nonlinear optical polymers," JOSA B, 15(1), 318-328 (1998).
[17] X. Zhang, B. Lee, C.-y. Lin, A. X. Wang, A. Hosseini, and R. T. Chen, "Highly Linear Broadband Optical Modulator Based on Electro-Optic Polymer," Photonics Journal, IEEE, 4(6), 2214-2228 (2012).
[18] X. Lin, T. Ling, H. Subbaraman, X. Zhang, K. Byun, L. J. Guo, and R. T. Chen, "Ultraviolet imprinting and aligned ink-jet printing for multilayer patterning of electro-optic polymer modulators," Optics letters, 38(10), 1597-1599 (2013).
[19] C.-Y. Lin, A. X. Wang, X. Zhang, B. S. Lee, and R. T. Chen, "EO-polymer waveguide based high dynamic range EM wave sensors," in SPIE OPTO, 2012, pp. 82580Y-82580Y-7.
[20] H. Subbaraman, X. Lin, T. Ling, X. Zhang, L. J. Guo, and R. T. Chen, "Printable EO Polymer Modulators," in CLEO: Science and Innovations, 2013, p. CW1O. 2.
[21] S. Huang, T.-D. Kim, J. Luo, S. K. Hau, Z. Shi, X.-H. Zhou, H.-L. Yip, and A. K.-Y. Jen, "Highly efficient electro-optic polymers through improved poling using a thin TiO 2-modified transparent electrode," Applied Physics Letters, 96(24), 243311-243311-3 (2010).
[22] X. Wang, C.-Y. Lin, S. Chakravarty, J. Luo, A. K.-Y. Jen, and R. T. Chen, "Effective in-device r33 of 735 pm/V on electro-optic polymer infiltrated silicon photonic crystal slot waveguides," Optics letters, 36(6), 882-884 (2011).
[23] X. Zhang, A. Hosseini, S. Chakravarty, J. Luo, A. K.-Y. Jen, and R. T. Chen, "Wide optical spectrum range, subvolt, compact modulator based on an electro-optic polymer refilled silicon slot photonic crystal waveguide," Optics letters, 38(22), 4931-4934 (2013).
[24] H. Chen, B. Chen, D. Huang, D. Jin, J. Luo, A.-Y. Jen, and R. Dinu, "Broadband electro-optic polymer modulators with high electro-optic activity and low poling induced optical loss," Applied Physics Letters, 93(4), 043507 (2008).
[25] T. Baehr-Jones, B. Penkov, J. Huang, P. Sullivan, J. Davies, J. Takayesu, J. Luo, T.-D. Kim, L. Dalton, and A. Jen, "Nonlinear polymer-clad silicon slot waveguide modulator with a half wave voltage of 0.25 V," Applied Physics Letters, 92(16), 163303 (2008).
[26] X. Zhang, A. Hosseini, H. Subbaraman, J. Luo, A. Jen, R. Chen, "Broadband Low-power Optical Modulator Based on Electro-optic Polymer Infiltrated Silicon Slot Photonic Crystal Waveguide," Frontiers in Optics/Laser Science Conference, Optical Society of America, 2014), p. FTu1D.4.
[27] X. Zhang, A. Hosseini, J. Luo, A. Jen, and R. Chen, "Ultralow Power Consumption of 1.5 nW Over Wide Optical Spectrum Range in Silicon Organic Hybrid Modulator," in CLEO: Science and Innovations, 2014, p. SM2G. 4.
[28] Z. Wang, N. Zhu, Y. Tang, L. Wosinski, D. Dai, and S. He, "Ultracompact low-loss coupler between strip and slot waveguides," Optics letters, 34(10), 1498-1500 (2009).
[29] C.-Y. Lin, X. Wang, S. Chakravarty, B. S. Lee, W. Lai, J. Luo, A. K.-Y. Jen, and R. T. Chen, "Electro-optic polymer infiltrated silicon photonic crystal slot waveguide modulator with 23 dB slow light enhancement," Applied Physics Letters, 97(9), 093304 (2010).
[30] J. Blasco, and C. Barrios, "Compact slot-waveguide/channel-waveguide mode-converter," in Lasers and Electro-Optics Europe, 2005. CLEO/Europe. 2005 Conference on, 2005, pp. 607-607.
[31] Y. Liu, T. Baehr-Jones, J. Li, A. Pomerene, and M. Hochberg, "Efficient Strip to Strip-Loaded Slot Mode Converter in Silicon-on-Insulator," Photonics Technology Letters, IEEE, 23(20), 1496-1498 (2011).
[32] R. Palmer, A. Luca, D. Korn, W. Heni, P. Schindler, J. Bolten, M. Karl, M. Waldow, T. Wahlbrink, and W. Freude, "Low-loss silicon strip-to-slot mode converters," IEEE Photonics Journal, (2013).
[33] W.-C. Lai, S. Chakravarty, X. Wang, C. Lin, and R. T. Chen, "On-chip methane sensing by near-IR absorption signatures in a photonic crystal slot waveguide," Optics letters, 36(6), 984-986 (2011).
[34] H. C. Nguyen, Y. Sakai, M. Shinkawa, N. Ishikura, and T. Baba, "10 Gb/s operation of photonic crystal silicon optical modulators," Optics Express, 19(14), 13000-13007 (2011).
[35] C. Caer, X. Le Roux, J. Oden, L. Vivien, N. Dubreuil, and E. Cassan, "Design and fabrication of hollow core slow light slot photonic crystal waveguides for nonlinear optics," in Asia Communications and Photonics Conference, 2013, p. AW4B. 1.
[36] A. Hosseini, X. Xu, D. N. Kwong, H. Subbaraman, W. Jiang, and R. T. Chen, "On the role of evanescent modes and group index tapering in slow light photonic crystal waveguide coupling efficiency," Applied Physics Letters, 98(3), 031107-031107-3 (2011).

[37]	A. Hosseini, X. Xu, H. Subbaraman, C.-Y. Lin, S. Rahimi, and R. T. Chen, "Large optical spectral range dispersion engineered silicon-based photonic crystal waveguide modulator," Opt. Express, 20(11), 12318-12325 (2012).

[38]	H. Subbaraman, X. Xu, J. Covey, and R. T. Chen, "Efficient light coupling into in-plane semiconductor nanomembrane photonic devices utilizing a sub-wavelength grating coupler," Optics Express, 20(18), 20659-20665 (2012).

[39]	X. Xu, H. Subbaraman, J. Covey, D. Kwong, A. Hosseini, and R. T. Chen, "Complementary metal–oxide–semiconductor compatible high efficiency subwavelength grating couplers for silicon integrated photonics," Applied Physics Letters, 101(3), 031109-031109-4 (2012).

[40]	J. Witzens, T. Baehr-Jones, and M. Hochberg, "Design of transmission line driven slot waveguide Mach-Zehnder interferometers and application to analog optical links," Optics Express, 18(16), 16902-16928 (2010).

[41]	R. Palmer, S. Koeber, D. L. Elder, M. Woessner, W. Heni, D. Korn, M. Lauermann, W. Bogaerts, L. Dalton, and W. Freude, "High-Speed, Low Drive-Voltage Silicon-Organic Hybrid Modulator Based on a Binary-Chromophore Electro-Optic Material," Journal of Lightwave Technology, 32(16), 2726-2734 (2014).

[42]	X. Zhang, A. Hosseini, J. Luo, A. K.-Y. Jen, and R. T. Chen, "Hybrid silicon-electro-optic-polymer integrated high-performance optical modulator," in SPIE Photonic West, OPTO, 2014, pp. 89910O-89910O-6.

[43]	X. Zhang, A. Hosseini, J. Luo, A. K.-Y. Jen, and R. T. Chen, [Ultraperformance Nanophotonic Modulator Based On Silicon Organic Hybrid Technology] IEEE, (2014).

[44]	R. Ding, T. Baehr-Jones, W.-J. Kim, B. Boyko, R. Bojko, A. Spott, A. Pomerene, C. Hill, W. Reinhardt, and M. Hochberg, "Low-loss asymmetric strip-loaded slot waveguides in silicon-on-insulator," Applied Physics Letters, 98(23), 233303 (2011).

[45]	A. Spott, T. Baehr-Jones, R. Ding, Y. Liu, R. Bojko, T. O'Malley, A. Pomerene, C. Hill, W. Reinhardt, and M. Hochberg, "Photolithographically fabricated low-loss asymmetric silicon slot waveguides," Optics Express, 19(11), 10950-10958 (2011).

[46]	F. Kish, "500Gb/s and Beyond PIC-Module Transmitters and Receivers," in Optical Fiber Communication Conference, 2014, p. W3I. 1.

[47]	D. Dai, J. Bauters, and J. E. Bowers, "Passive technologies for future large-scale photonic integrated circuits on silicon: polarization handling, light non-reciprocity and loss reduction," Light: Science & Applications, 1(3), e1 (2012).

[48]	Z. Yuan, A. Anopchenko, N. Daldosso, R. Guider, D. Navarro-Urrios, A. Pitanti, R. Spano, and L. Pavesi, "Silicon nanocrystals as an enabling material for silicon photonics," Proceedings of the IEEE, 97(7), 1250-1268 (2009).

[49]	B. G. Lee, A. V. Rylyakov, W. M. Green, S. Assefa, C. W. Baks, R. Rimolo-Donadio, D. M. Kuchta, M. H. Khater, T. Barwicz, and C. Reinholm, "Monolithic Silicon Integration of Scaled Photonic Switch Fabrics, CMOS Logic, and Device Driver Circuits," Journal of Lightwave Technology, 32(4), 743-751 (2014).

[50]	X. Zheng, and A. V. Krishnamoorthy, "Si photonics technology for future optical interconnection," in SPIE/OSA/IEEE Asia Communications and Photonics, 2011, pp. 83091V-83091V-11.

[51]	A. V. Krishnamoorthy, K. W. Goossen, W. Jan, X. Zheng, R. Ho, G. Li, R. Rozier, F. Liu, D. Patil, and J. Lexau, "Progress in low-power switched optical interconnects," Selected Topics in Quantum Electronics, IEEE Journal of, 17(2), 357-376 (2011).

[52]	F. E. Doany, C. L. Schow, C. W. Baks, D. M. Kuchta, P. Pepeljugoski, L. Schares, R. Budd, F. Libsch, R. Dangel, and F. Horst, "160 Gb/s bidirectional polymer-waveguide board-level optical interconnects using CMOS-based transceivers," Advanced Packaging, IEEE Transactions on, 32(2), 345-359 (2009).

[53]	C.-H. Chen, C. Li, A. Shafik, M. Fiorentino, P. Chiang, S. Palermo, and R. Beausoleil, "A WDM Silicon Photonic Transmitter Based on Carrier-Injection Microring Modulators," (2014).

[54]	R. Ryf, S. Randel, N. K. Fontaine, M. Montoliu, E. Burrows, S. Chandrasekhar, A. H. Gnauck, C. Xie, R.-J. Essiambre, and P. Winzer, "32-bit/s/Hz spectral efficiency WDM transmission over 177-km few-mode fiber," in Optical Fiber Communication Conference, 2013, p. PDP5A. 1.

[55]	W. M. Green, M. J. Rooks, L. Sekaric, and Y. A. Vlasov, "Ultra-compact, low RF power, 10 Gb/s silicon Mach-Zehnder modulator," Optics Express, 15(25), 17106-17113 (2007).

[56]	J. Ding, R. Ji, L. Zhang, and L. Yang, "Electro-optical response analysis of a 40 Gb/s silicon Mach-Zehnder optical modulator," Journal of Lightwave Technology, 31(14), 2434-2440 (2013).

[57]	D. M. Gill, J. E. Proesel, C. Xiong, J. Rosenberg, M. Khater, T. Barwicz, S. Assefa, S. M. Shank, C. Reinholm, and E. Kiewra, "Monolithic Travelling-Wave Mach-Zehnder Transmitter with High-Swing

Stacked CMOS Driver," in CLEO: Science and Innovations, 2014, p. SM2G. 3.
[58] C. DeRose, "Integrated RF Silicon Photonics from High Power Photodiodes to Linear Modulators," in Integrated Photonics Research, Silicon and Nanophotonics, 2014, p. IW2A. 1.
[59] D. Mahgerefteh, [Transmission system comprising a semiconductor laser and a fiber grating discriminator] Google Patents, (2000).
[60] X. Zheng, E. Chang, P. Amberg, I. Shubin, J. Lexau, F. Liu, H. Thacker, S. S. Djordjevic, S. Lin, and Y. Luo, "A high-speed, tunable silicon photonic ring modulator integrated with ultra-efficient active wavelength control," Optics Express, 22(10), 12628-12633 (2014).
[61] C.-Y. Lin, A. X. Wang, B. S. Lee, X. Zhang, and R. T. Chen, "High dynamic range electric field sensor for electromagnetic pulse detection," Optics Express, 19, 17372-17377 (2011).
[62] A. B. Matsko, A. A. Savchenkov, V. S. Ilchenko, D. Seidel, and L. Maleki, "On the sensitivity of all-dielectric microwave photonic receivers," Journal of Lightwave Technology, 28(23), 3427-3438 (2010).
[63] R. Chang, V. Lomakin, and E. Michielssen, "Coupling electromagnetics with micromagnetics," in Antennas and Propagation Society International Symposium (APSURSI), 2012 IEEE, 2012, pp. 1-2.
[64] X. Zhang, A. Hosseini, H. Subbaraman, S. Wang, Q. Zhan, J. Luo, A. K. Jen, and R. Chen, "Wideband Electromagnetic Wave Sensing Using Electro-optic Polymer Infiltrated Silicon Slot Photonic Crystal Waveguide," in CLEO: Science and Innovations, 2014, p. SM2M. 5.